\documentclass[final,3p,times,twocolumn,pdftex]{elsarticle}
\biboptions{comma,compress}
\usepackage{macros}
\usepackage{amssymb,amsmath}
\usepackage{mathrsfs}
\usepackage{booktabs,dcolumn}
\begin{document}
%
%
\begin{frontmatter}

\title{%
{%
\vspace{-3.0cm}
\small\hfill\parbox{32.0mm}{\raggedleft%
MS-TP-12-13\\
LPT-Orsay/12-101\\
CERN-PH-TH/2012-256\\
HU-EP-12/31\\
TCDMATH 12-08\\
DESY 12-167\\
SFB/CPP-12-74
}}\\[0.5cm]
B-physics from non-perturbatively renormalized HQET\\ 
in two-flavour lattice QCD%
\\[1.25em]
\vbox{\centerline{%
\includegraphics[width=2.5cm]{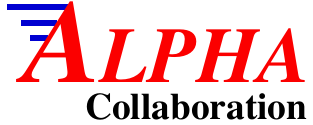}
}}
}

\author[label1]{Fabio Bernardoni} 
\address[label1]{%
NIC, DESY,
Platanenallee 6, D-15738 Zeuthen, Germany
}
\author[label2]{Beno\^it Blossier}
\address[label2]{%
CNRS et Universit\'e Paris-Sud XI, 
Laboratoire de Physique Th\'eorique,
B\^atiment 210, F-91405 Orsay Cedex, France
}
\author[label3]{John Bulava}
\address[label3]{%
CERN, 
Physics Department, TH Unit,
CH-1211 Geneva 23, Switzerland
}
\author[label4]{Michele Della Morte}
\address[label4]{%
Johannes Gutenberg Universit\"at Mainz,
Institut f\"ur Kernphysik,
Becher Weg 45, D-55099 Mainz, Germany
}
\author[label5]{Patrick Fritzsch}
\address[label5]{%
Humboldt Universit\"at Berlin, 
Institut f\"ur Physik,
Newtonstra{\ss}e 15, D-14289 Berlin, Germany
}
\author[label6]{Nicolas Garron}
\address[label6]{%
Trinity College, School of Mathematics, Dublin 2, Ireland
}
\author[label2]{Antoine G\'erardin}
\author[label7]{%
Jochen Heitger\corref{cor1}}
\address[label7]{%
Westf\"alische Wilhelms-Universit\"at M\"unster, 
Institut f\"ur Theoretische Physik,
Wilhelm-Klemm-Stra{\ss}e 9, D-48149 M\"unster, Germany
}
\cortext[cor1]{%
Speaker at
{\it QCD 2012, Montpellier, France, 2 -- 6 July 2012}}
\ead{heitger@uni-muenster.de}
\author[label4]{Georg M. von Hippel}
\author[label1]{Hubert Simma}

\begin{abstract}
\noindent
We report on the ALPHA Collaboration's lattice B-physics programme based on 
$N_{\rm f}=2$ O($a$) improved Wilson fermions and HQET, including all NLO 
effects in the inverse heavy quark mass, as well as non-perturbative 
renormalization and matching, to fix the parameters of the effective theory. 
Our simulations in large physical volume cover 3 lattice spacings
$a\approx (0.08-0.05)\,\Fm$ and pion masses down to $190\,\MeV$ to control 
continuum and chiral extrapolations. 
We present the status of results for the b-quark mass and the 
${\rm B}_{({\rm s})}$-meson decay constants, $\fB$ and $\fBs$.
\end{abstract}

% ...

\journal{Nucl. Phys. B Proc. Suppl.}

\end{frontmatter}
%
%
\section{B-physics and lattice QCD}
\label{Sec_Bphys}
\begin{figure}[htb]
\begin{center}
\includegraphics[width=0.475\textwidth]{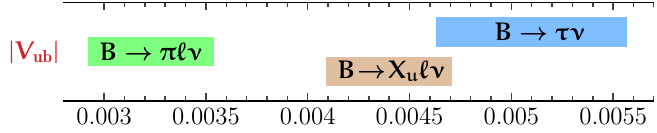}
\caption{\sl%
Observed tension among different 
$|V_{\rm ub}|$--determinations~\cite{PDBook};
$\pm 1\,\sigma$ bands are shown.
}\label{fig:vubpuzzle2012}
\end{center}
\vspace{-0.25cm}
\end{figure}

\noindent
Lattice simulations of QCD have established as a sound tool to compute 
strong interaction effects for accurate phenomenology in heavy flavour 
physics. 
For B-meson weak decays, which constrain the CKM Unitarity Triangle, 
lattice QCD results for the involved low-energy hadronic matrix elements 
in conjunction with experimental studies decisively contribute to stringent 
tests of the self-consistency of the Standard Model and complement direct 
searches for New Physics.
Since the significance of such precision tests in the beauty sector is
predominantly limited by theoretical uncertainties, lattice computations 
with an overall accuracy of a few \% are highly desirable.
Let us highlight the "$V_{\rm ub}$--puzzle" illustrated in 
Fig.~\ref{fig:vubpuzzle2012}, which has drawn the community's attention in 
the recent past.
%
%
$|V_{\rm ub}|$ can be determined from inclusive semi-leptonic processes
${\rmB\to X_{\rm u}\ell\nu}$, from exclusive semi-leptonic 
${\rmB\to\pi\ell\nu}$ decays and from the leptonic one, ${\rmB\to\tau\nu}$.
In the latter two, the hadronic form factor $f_{+}(q^2)$ and the B-meson
decay constant $\Fb$ enter, respectively, so that lattice QCD input is 
required to extract $|V_{\rm ub}|$. 
As currently there is a $\sim 3\sigma$ tension between its two exclusive 
(semi-leptonic and leptonic) determinations\footnote{%
At the ICHEP 2012 Conference, the Belle Collaboration reported a new 
result for 
${\mathscr B}(\rmB\to\tau\nu)$~\cite{Btaunu:BELLE_ichep12,Btaunu:BELLE},
obtained on basis of a new data set using a more sophisticated tagging
of the $\rmB$.
Taken alone, this would yield a value for $|V_{\rm ub}|$ compatible with the
exclusive semi-leptonic determination.
However, this result has not yet been confirmed by other collaborations,
and more data and a careful study of all systematics are still required
before drawing final conclusions.
},
as well as an inconsistency with the estimate from inclusive decays, 
precision lattice QCD calculations can contribute to resolve this 
tension~\cite{Lunghi:2010gv,lat11:ckmphys,Lenz:2012az}.
\section{Non-perturbative HQET}
\label{Sec_npHQET}
\noindent
The particular challenge of B-physics on the lattice lies in the many
disparate scales, 
ranging from its inverse extent
(as IR cutoff) 
over the hierarchy of differently flavoured hadron masses up to the inverse 
of the lattice spacing $a$
(as UV cutoff),  
to be treated simultaneously in the numerical simulations.
Since lattice sizes that are computationally manageable today have 
$am_{\rm b}>1$, discretization effects get most severe for heavy quark 
systems with b-quarks and escape brute force simulations.
Our approach to lattice B-physics is therefore based on the Heavy Quark 
Effective Theory (HQET) for the 
b-quark~\cite{stat:eichhill1,stat:eichhill_1m}, which consists in a 
systematic expansion of its QCD action and heavy-light correlation functions 
in $\lQCD/\mb\ll 1$ around the static limit ($\mb\to\infty$).

The Lagrangian entering the heavy quark field's lattice action
${S}_{\rm HQET}=a^4{\T \sum_x}\,{\mathscr L}_{\rm HQET}(x)$ in HQET at NLO, 
i.e., including $\Or(1/\mb)$ terms, reads:
\be
{\mathscr L}_{\rm HQET}=
\heavyb D_{0}\heavy-\omkin\Okin-\omspin\Ospin\,,
\label{lhqet}
\ee
with $\heavy$ satisfying $P_+\heavy=\heavy$, $P_+={{1+\gamma_0}\over{2}}$.
The parameters $\omega_{\rm kin}$ and $\omega_{\rm spin}$ are formally 
$\Or(1/\mb)$ and multiply the dimension--5 operators
$\Okin=\heavyb\vecD^2\heavy$ and $\Ospin=\heavyb\vecsigma\vecB\,\heavy$,
representing interaction terms due to the motion and the spin of the 
heavy quark.
Thus, ${S}_{\rm HQET}$ has $\Or(\lQCD^2/\mb^2)$ truncation errors, and lattice
artifacts only scale as $(a\lQCD)^n$ rather than $(a\mb)^n$.
Analogously, local composite fields are introduced in the effective lattice
theory.
For instance, the NLO HQET expansion of the zero-momentum projected time 
component of the heavy-light axial vector current can be written as
\bea
\hspace{-0.25cm}
\Arenhqet(x_0)
& = &
\zahqet a^3{\T \sum_{\vecx}}\left[\Astat(x)+{\cah{1}}\Ah{1}(x)\right]\,,
\nonumber\\
\hspace{-0.25cm}
\Astat(x)
& = &
\lightb(x)\,\gamma_0\gamma_5\,\heavy(x)\,,
\nonumber\\
\hspace{-0.25cm}
\Ah{1}(x)
& = & 
\lightb(x)\,\gamma_5\gamma_i\,\half\,
\big(\nabsym{i}-\lnabsym{i}\,\;\big)\,\heavy(x)\,,
\label{ahqet}
\eea
$\nabsym{i}\;$ being the spatial covariant derivative.
The relation 
$\fps\sqrt{\mps}=\ketbra{\,0\,}{\,A_{0,{\rm R}}(0)\,}{\,{\rm PS}({\bf p}=0)\,}$ 
to the pseudoscalar decay constant will be used to calculate 
$f_{{\rm B}_{({\rm s})}}$ below.

HQET treats the $\Or(1/\mb)$ interactions terms in (\ref{lhqet}) as local 
space-time insertions in static correlations functions.
For correlators of some multi-local fields $\op{}$ and up to 
$1/m_{\rm b}$--corrections to the operator itself (irrelevant when spectral 
quantities are considered), this means
\bea
\langle\op{}\rangle=
\langle\op{}\rangle_{\mrm{stat}}+a^4\sum_x\Big\{\,
& &
\hspace{-0.75cm}
\,\omkin\langle\op{}\Okin(x)\rangle_{\mrm{stat}}
\label{ev}
\\[-2ex]
& &
\hspace{-0.75cm}
+\,\omspin\langle\op{}\Ospin(x)\rangle_{\mrm{stat}}
\,\Big\}\,,
\nonumber
\eea
where $\langle\op{}\rangle_{\rm stat}$ is the expectation value in the static 
theory.

Still, for lattice HQET applications to lead to precise and controlled
results, two issues had to be solved.

1.) The exponential growth of the noise-to-signal ratio in static-light 
correlation functions with Euclidean time, caused by the linear divergence 
in the binding energy $E^{\rm stat}$ of the static-light system, which is
particularly severe for the Eichten-Hill action~\cite{stat:eichhill_za}.
This is overcome by so-called "HYP-smeared"~\cite{HYP:HK01} discretizations 
of the static quark action, improving the statistical precision of the 
correlators substantially~\cite{fbstat:pap1,HQET:statprec}.

2.) Operator mixing in the effective theory induces UV power divergences in 
the lattice spacing that must be subtracted non-perturbatively:
The formal definition of lattice HQET and its composite fields in 
\eqref{lhqet} and \eqref{ahqet} involves the (a priori free) effective 
couplings
\be
\boldsymbol{\omega}\equiv
\Big\{\,m_{\rm bare},\ln\zahqet,\cah{1},\omkin,\omspin\,\Big\}\,.
\label{hqetparam}
\ee
Here, the energy shift $m_{\rm bare}$ is an additive mass renormalization.
It absorbs the $1/a$--divergence of the static energy, $E^{\rm stat}$,
and a $1/a^2$--divergence at $\Or(1/\mb)$. 
Hence, a phenomenologically relevant predictive power of lattice HQET 
is only guaranteed, once these 
\emph{HQET parameters $\boldsymbol{\omega}=\{\omega_i\}$ have been fixed 
non-perturbatively} such that no uncancelled power divergences in $1/a$, 
which would remain in perturbation theory~\cite{Maiani:1992az}, can preclude
to take the continuum limit.

A solution to 2.) was developed in~\cite{HQET:pap1} and relies upon a 
\emph{non-perturbative matching of HQET and QCD in finite volume}.
The implementation of this strategy by our collaboration has led to NLO HQET
computations of the b-quark mass, B-meson spectroscopy and decay constants 
in the quenched approximation 
($\nf=0$)~\cite{HQET:mb1m,HQET:param1m,HQET:msplit,HQET:fb1m},
as well as in the more realistic two-flavour theory 
\cite{lat08:hqettests,impr:babp_nf2,lat11:hqetNf2,HQET:Nf2param1m}, of which
we give an overview in the following.
\subsection{General strategy}
\label{Sec_npHQET_strat}
\begin{figure}[htb]
\begin{center}
\includegraphics[width=0.475\textwidth]{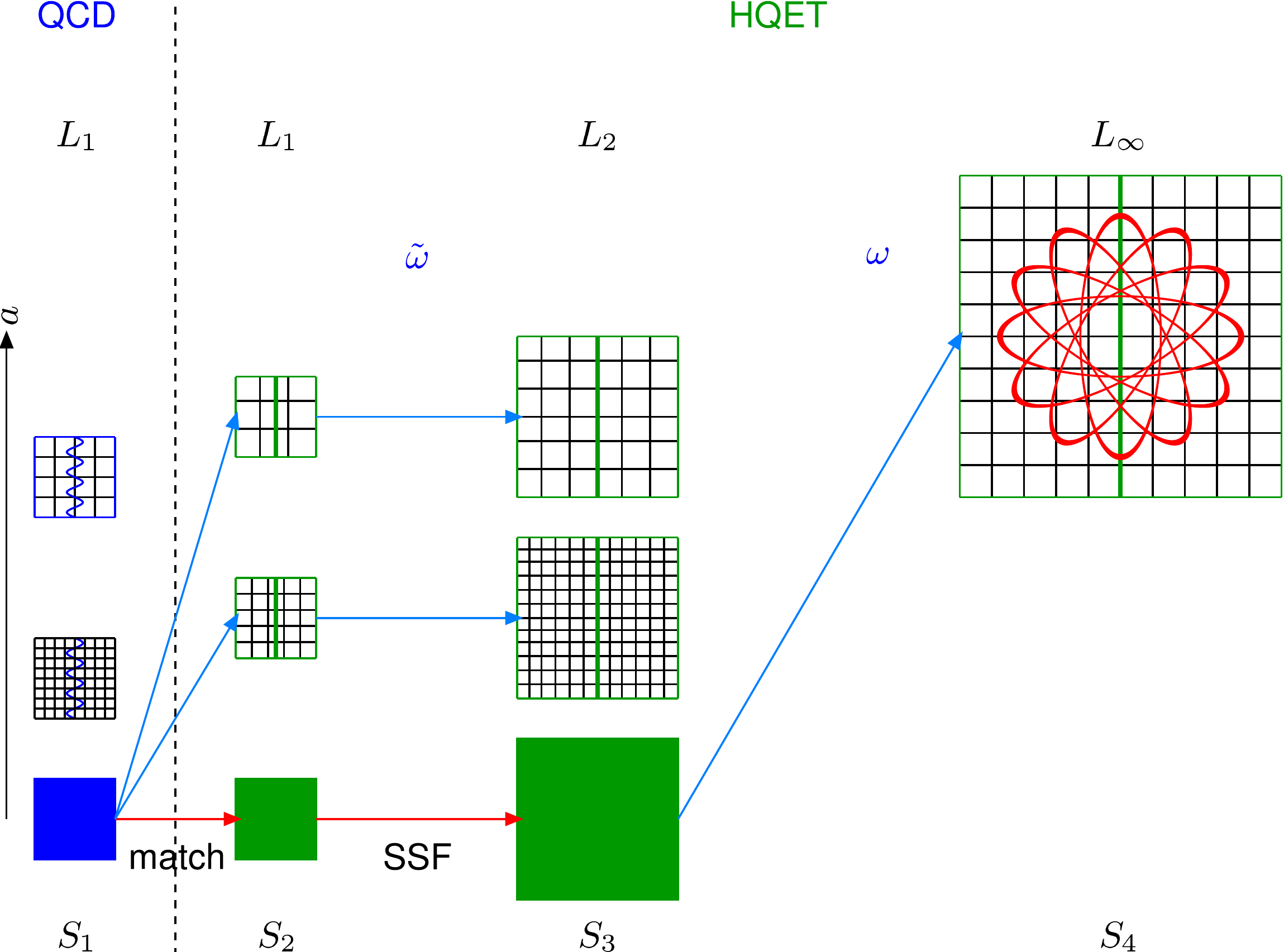}
\caption{\sl%
Idea of lattice HQET computations for B-physics phenomenology via a 
non-perturbative determination of HQET parameters from small-volume QCD 
simulations.
The step scaling method makes contact to physically large volumes 
$L_{\infty}$.
The whole construction is such that the continuum limit can be taken for 
all pieces. 
}\label{fig:strat}
\end{center}
\end{figure}

\noindent
The computational strategy of our 
approach~\cite{HQET:pap1,HQET:param1m,HQET:Nf2param1m}, in which matching 
and renormalization are performed simultaneously \emph{and} 
non-perturbatively\footnote{%
As soon as $1/\mb$--corrections are included, matching must be done 
non-perturbatively in order not to spoil the asymptotic convergence of the 
series.
Otherwise, the perturbative truncation error from the matching coefficient 
of the static term becomes much larger than the power corrections 
$\sim\lQCD/\mb$ of HQET, as $\mb\to\infty$.
},
is sketched in Fig.~\ref{fig:strat}.
%
%
The matching part is performed in a small volume of $L_1\approx 0.4\,\Fm$, 
where owing to $a\mb\ll 1$ numerical simulations with a relativistic b-quark 
are feasible.
The bare HQET parameters $\omega_i$ of the Lagrangian and the time component 
of the heavy-light axial current are fixed by imposing matching conditions 
$\PhiHQET(z,a)\stackrel{!}{=}\PhiQCD(z,0)=
\lim_{a\to 0}\PhiQCD(z,a)$.
The quark mass dependence of (non-perturbatively renormalized) QCD is 
inherited by the HQET parameters $\omega_i$.
It enters through the dimensionless variable $z\equiv L_1M$, where $M$ is 
the renormalization group invariant (RGI) mass~\cite{impr:babp_nf2}.
Then a recursive finite-size scaling step $L_1\to L_2=2L_1$ is used to 
reach larger volumes and lattice spacings $a$, by which connection with 
phenomenology in $L_{\infty}\gtrsim \max(2\,\Fm,4/\mpi)$ can be made. 
As a result of~\cite{HQET:Nf2param1m}, the HQET parameters~\eqref{hqetparam},
$\boldsymbol{\omega}=\boldsymbol{\omega}(z,a)$, absorbing the logarithmic 
and power divergences of HQET, are now non-perturbatively known for 
renormalized QCD quark masses from the charm to beyond the beauty region 
(parameterised by $z\in\left\{4,6,7,9,11,13,15,18,21\right\}$) and for $a$'s 
corresponding to the bare gauge couplings of the available two-flavour 
configuration ensembles in large volume used to compute B-physics 
observables.
\subsection{Large-volume computations and techniques}
\label{Sec_npHQET_largeV}
\noindent
Our large-volume gauge configuration ensembles are characterised by the
plaquette gauge action and a sea of $\nf=2$ mass-degenerate 
non-perturbatively $\Or(a)$ improved dynamical Wilson quarks.
To be able to extrapolate to the physical pion mass, several pseudoscalar 
(sea) masses in the range $(190\lesssim m_{\rm PS}\lesssim 440)\,\MeV$ with
$L\,m_{\rm PS}\gtrsim 4$ are considered such that finite-volume effects are
expected to be negligible.
Moreover, the configurations cover 3 $\beta$--values $\{5.2,5.3,5.5\}$
with lattice spacings $a\in\{0.08\,\Fm,0.07\,\Fm,0.05\,\Fm\}$ 
\cite{scale:fk_Nf2} to control the extrapolation to the continuum limit.

For the numerical simulations to generate these two-flavour QCD
configurations, we employ M.~L\"uscher's implementation of the Hybrid Monte 
Carlo (HMC) algorithm with domain decomposition~\cite{ddhmc:luescher3}
and an adaption~\cite{lat10:marina}, which combines the deflated SAP 
solver~\cite{ddhmc:luescher2,deflat:luescher1,deflat:luescher2} with mass 
preconditioning~\cite{hmc:hasenb1}, chronological 
inversion~\cite{Brower:1995vx} and multiple time scale 
integration~\cite{hmc:mtsi1,hmc:mtsi2}.
Large trajectory length~\cite{Nf2SF:autocorr,hmc:slowingdown} and long runs 
are used to ensure that our ensembles are not biased by the critical slowing 
down of the QCD simulations.
For a careful and conservative error estimation, a binned jackknife 
procedure is applied (which is being cross-checked by the method advocated 
in~\cite{hmc:slowingdown}).
All large-volume configuration ensembles have been produced and are shared 
within the Coordinated Lattice Simulations (CLS) effort by several lattice 
QCD teams in Europe~\cite{community:CLS}.

Our determination of the B-meson spectrum and decay constants in two-flavour
QCD is based on their HQET expansions in terms of the known HQET parameters 
and associated HQET energies and matrix elements at the static and  
$1/m_{\rm b}$--order.
The latter are extracted from static-light correlation functions evaluated 
on the available large-volume CLS ensembles by solving the Generalised 
Eigenvalue Problem (GEVP) discussed and applied 
in~\cite{gevp:LW,HQET:gevp,HQET:msplit,HQET:fb1m}, allowing for a better
control of excited state contaminations of the correlators.
More specifically, the GEVP analysis amounts to compute a 
$(N\times N)$--matrix of correlators with the desired static and 
$\Or(1/m_{\rm b})$ insertions, where each entry of the matrix corresponds to 
a different Gaussian smearing level~\cite{smear:wupp} of the light quark 
field in the B-meson interpolating quark bilinear.
In these computations, variance reduction in the light quark sector is 
achieved by stochastic all-to-all propagators (with 8 noise sources and full 
time-dilution) \cite{ata:dublin,HQET:msplit}, while for the static quark 
propagators, we use two variants of the HYP-smeared static actions, 
HYP1/2~\cite{fbstat:pap1,HQET:statprec}, already mentioned in 
Sect.~\ref{Sec_npHQET}.
Solving the GEVP numerically gives rise to new estimators for effective 
energies and hadron-to-vacuum matrix elements, which converge faster with 
Euclidean time separation than standard ratios, since a larger gap governs
the excited state corrections.
I.e., corrections to ground state energies and matrix elements fall off in 
$t$ and $t_0$ as  
$\,\propto\exp\left\{-(E_{N+1}-E_1)\,t\right\}$
and 
$\,\propto\exp\left\{-(E_{N+1}-E_1)\,t_0\right\}
\times\exp\left\{-(E_2-E_1)\,(t-t_0)\right\}$,
respectively, where $t_0<t<2t_0$ and $N$ labels the $N^{\rm th}$ excited 
state (and $N=3$ in practice). 
Our final estimates are then obtained as plateau averages over ranges 
conservatively chosen by varying $t_{\rm min}$ for fixed $t_{\rm max}$ such that 
$\Or\big(\Exp^{-(E_{N+1}-E_1)t}\big)\sim
\sigma_{\rm sys}\lesssim\frac{1}{3}\,\sigma_{\rm stat}$,
thereby minimising our systematic errors.
For more details we refer 
to~\cite{HQET:gevp,HQET:msplit,HQET:fb1m,lat12prlm:hqetNf2,HQET:Nf2}.
\section{Results}
\label{Sec_res}
\noindent
In this section we summarise the status of results of our $\nf=2$ B-physics 
project, as it was reported at summer conferences in 2012, see 
also~\cite{lat12prlm:hqetNf2,ichep12prlm:hqetNf2}.
A final account of our work will be given later, once the full statistics
of all CLS ensembles has been analysed~\cite{HQET:Nf2}.

Our HQET energies and matrix elements extracted from static-light 
correlators split into two sets, one for the B-meson sector with the valence
quark masses set equal to the CLS sea quark mass values, and another one for 
the $\Bs$-meson sector with the valence quark tuned to the physical strange 
quark~\cite{scale:fk_Nf2}, corresponding to a partially quenched setup.
%
\begin{figure}[htb]
\begin{center}
\includegraphics[trim=0 0 250 0,clip=true,width=0.485\textwidth]{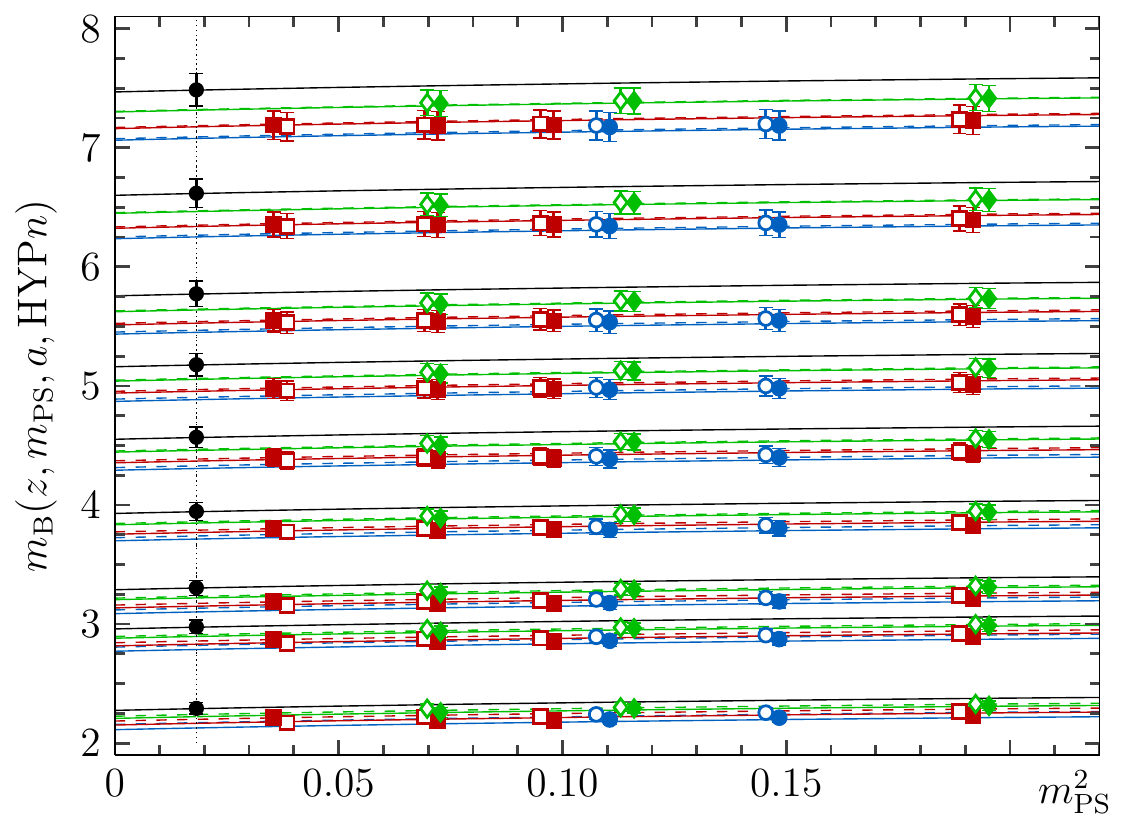}
\caption{\sl%
Joint chiral and continuum extrapolation of the heavy-light
pseudoscalar meson mass (\ref{hqetexpan_mB}) in NLO HQET to the
ansatz (\ref{fitansatz_mB}) for fixed $z$.
Recall that the $z$--dependence originates from the initial matching step 
to finite-volume QCD determining the HQET parameters.
(Blue, red and green points refer to $\beta=5.2$, $5.3$ and $5.5$,
respectively, while filled/open symbols belong to the HYP1/2 static 
actions.)
}\label{fig:Mb_LOxPT}
\end{center}
\end{figure}
%
\subsection{The b-quark's mass}
\label{Sec_res_mb}
\noindent
To begin with, we apply the non-perturbative results on the HQET parameters 
from the matching step together with the HQET energies from the large-volume 
GEVP analysis of the CLS ensembles to calculate the b-quark mass.
To this end one writes down the NLO HQET expansion (i.e., to first order in 
$1/\mb$) of the heavy-light pseudoscalar meson mass as 
\be
\mB=
m_{\rm bare}+E^{\rm stat}
+\omega_{\rm kin}E^{\rm kin}
+\omega_{\rm spin}E^{\rm spin}\,.
\label{hqetexpan_mB}
\ee
Beside the dependence on the light pseudoscalar (sea) mass $m_{\rm PS}$ and 
the lattice spacing $a$ of the CLS configurations, we also have to account 
for the apparent heavy quark mass ($z$) dependence of the HQET parameters.
In fact, this $z$--dependence can now be exploited to fix the HQET
parameters $\omega_i=\omega_i(z,a)$ for once by imposing the condition
\be
\left.\mB(z,m_{\pi},a)\,\right|_{\,z=z_{\rm b}}\equiv
\mB^{\rm exp}=5279.5\,\MeV
\quad\mbox{\cite{PDBook}}
\label{fixeq_mb}
\ee
in the continuum, which \emph{defines} the physical value of the b-quark
mass at NLO of HQET. 

For given $z$, the l.h.s. is obtained by evaluating \eqref{hqetexpan_mB} 
for each value of the lattice spacing and sea quark mass of the CLS 
ensembles, followed by a global fit of $\mB$ (simultaneously for two 
variants of the aforementioned HYP-smeared actions, ${\rm HYP}n$, $n=1,2$) 
to the ansatz for a combined chiral 
($m_{\rm PS}\to m_{\pi}\equiv m_{\pi}^{\rm exp}=135\,\MeV$~\cite{PDBook}) and 
continuum ($a\to 0$) extrapolation,
\bea
& &
\hspace{-0.5cm}
\mB(z,m_{\rm PS},a,{\rm HYP}n)=
\nonumber\\
& &
\hspace{-0.5cm}
B(z)+Cm^2_{\rm PS}-\frac{3\,\hat{g}^2}{16\pi f_{\pi}^2}\,m^3_{\rm PS} 
+D_{{\rm HYP}n}\,a^2\,,
\label{fitansatz_mB}
\eea
with $\fpi\equiv f_{\pi}^{\,\rm exp}=130.4\,\MeV$~\cite{PDBook} and 
$\hat{g}=0.51(2)$~\cite{lat10:michael} the $\Bstar\rmB\pi$--coupling in the 
static approximation for the b-quark.
These extrapolations to the physical point are shown in 
Fig.~\ref{fig:Mb_LOxPT}.

%
\begin{figure}[htb]
\begin{center}
\includegraphics[trim=325 0 0 0,clip=true,width=0.45\textwidth]{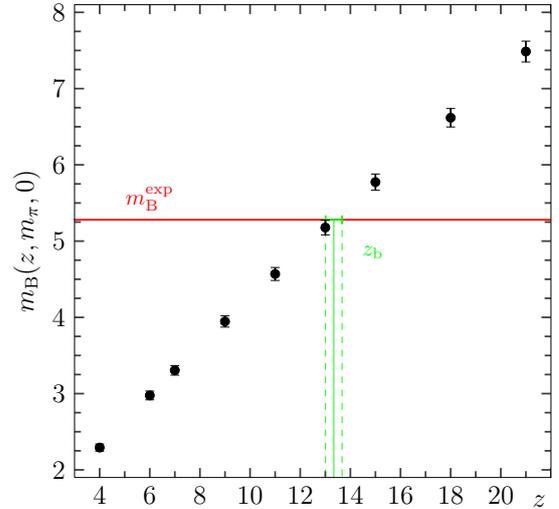}
\caption{\sl%
$z$--dependence of $\mB$ in the continuum limit and graphical solution 
of~(\ref{fixeq_mb}), which determines the physical b-quark mass $\zb$.
($z=L_1M$ denotes the dimensionless RGI heavy quark mass.)
}\label{fig:Mb_interpol}
\end{center}
\end{figure}
%
The solution of (\ref{fixeq_mb}) for $a=0$ as the physical condition 
defining the b-quark mass yields the dimensionless RGI b-quark mass
$\zb\equiv L_1\Mb$ and is illustrated in Fig.~\ref{fig:Mb_interpol}.
Converting $L_1$ to physical units, achieved via setting the lattice scale
through $\fk$ in~\cite{scale:fk_Nf2}, and translating with 4--/3--loop RG
running of the coupling/mass to the conventional $\MSbar$ scheme, we obtain 
as our result for the b-quark's mass in HQET at $\Or(1/\mb)$ in the $\nf=2$ 
theory presented at this year's summer conferences:
\bea
& &
z_{\rm b}=13.34(33)(13)_{z}
\nonumber\\
\Leftrightarrow
& &
\mbbMS(\mbbar)=4.22(10)(4)_{z}\,\GeV\,.
\label{res_mb}
\eea
The first error covers all statistical and systematic errors, including
those from the GEVP analysis, the various extrapolations and the scale
setting, while the second uncertainty of about 1\% stems from the quark mass 
renormalization in QCD, entering the finite-volume matching 
step~\cite{impr:babp_nf2,HQET:Nf2param1m}, and has to be added in quadrature.
Since we find the difference of the NLO HQET result (\ref{res_mb}) and the
corresponding number in the static approximation (LO HQET) to be very small,
we conclude that the truncation error of $\Or(\lQCD^3/\mb^2)$ to 
(\ref{res_mb}) in the HQET expansion is negligible compared to our overall 
error.

Our result (\ref{res_mb}) also compares very well with other recent 
determinations and the value quoted by the Particle Data Group,
see the (incomplete) compilation in Tab.~\ref{tab:mb}.
Note that some determinations claim very small errors, although they are 
based on perturbation theory or lattice data with heavy quark masses in
lattice units close to 1.
%
\begin{table}[htb]
\begin{center}
\small
\renewcommand{\arraystretch}{1.25}
\setlength{\tabcolsep}{3.5pt}
\begin{tabular}{lll} 
\toprule
$\mbbMS(\mbbar)/\GeV$ & remarks, method & ref. \\
\midrule
4.347(48) & lattice, $\nf=0$, NLO HQET & \cite{HQET:mb1m} \\
4.22(11)  & lattice, $\nf=2$, NLO HQET & eq.~(\ref{res_mb}) \\
4.29(14)  & lattice, $\nf=2$, extrapolation & \cite{mbottom:etmc} \\
4.164(23) & lattice, $\nf=3$, extrapolation & \cite{mbottom:hpqcd} \\
4.163(16) & perturbation theory \& data & \cite{mbottom:karlsruhe} \\
4.236(69) & perturbation theory \& QCD inputs & \cite{mbottom:narison_qcd12extd} \\
4.18(3)   & PDG average 2012 & \cite{PDBook} \\
\bottomrule
\end{tabular}
\caption{\sl%
Compilation of some recent determinations of $\mb$.
As for the lattice results \cite{mbottom:etmc} and \cite{mbottom:hpqcd},
the former uses extrapolations of relativistic data around the charm
to known static limits, while the latter employs moments of current-current
correlators with Highly Improved Staggered Quarks (HISQ)~\cite{hisq:hpqcd} 
extrapolated to the b-scale.
\cite{mbottom:karlsruhe} and \cite{mbottom:narison_qcd12extd} rely on
QCD sum rules.
For more details, see the cited references.
}\label{tab:mb}
\end{center}
\end{table}
%

After the determination of the value of the physical b-quark mass (and thus 
$\zb$) from $\mB$, we can fix the HQET parameters to 
$\omega_i(a)=\omega_i(\zb,a)$ and use those in any successive HQET 
computation of B-physics obervables.
\subsection{B-meson decay constants}
\label{Sec_res_fb}
\noindent
To determine the B-meson decay constants $\fB$ and $\fBs$, we now combine 
the HQET parameters with the matrix elements resulting from the GEVP 
analysis.
Distinguishing the heavy-light ${\rm B}_{\rm r}$--meson decay constants,
where the light flavour can be either a valence ($=$ sea) quark flavour 
${\rm r}=\{{\rm u,d}\}\equiv{\rm l}$ or a valence strange one, 
${\rm r}={\rm s}$, their NLO HQET expansions in terms of the HQET 
parameters $\omega_i$ read
\bea
\hspace{-0.75cm}
\fBq{r}\sqrt{{\T \frac{\mBq{r}}{2}}}
& = &
\zahqet\left(1+b^{{\rm stat}}_{\rm A}am_{\rm q,r}\right)p^{\rm stat}_{\rm r}
\label{hqetexpan_fB}
\\
&   &
\times\,\left(
1+\omega_{\rm kin}p^{\rm kin}_{\rm r}+\omega_{\rm spin}p^{\rm spin}_{\rm r}
+\cah{1}p^{\rm A^{(1)}}_{\rm r}
\right)\,.
\nonumber
\eea
As explained before, the $\omega_i$ are understood to be taken at the 
physical b-quark mass, $z_{\rm b}$, and the $p^{\rm X}$, 
${\rm X}\in\{{\rm stat},{\rm kin},{\rm spin},{\rm A^{(1)}}\}$, denote the 
previously extracted GEVP plateau values of the associated effective matrix 
elements of Sect.~\ref{Sec_npHQET_largeV}.
The improvement coefficient $b^{{\rm stat}}_{\rm A}$ is known to 1--loop
perturbation theory\footnote{%
For the subtracted bare quark masses appearing here, the additive 
improvement term $b^{{\rm stat}}_{\rm A}am_{\rm q,r}$ is numerically very small 
in practice.
} 
from~\cite{castat:HYP}.

\begin{figure}[htb]
\begin{center}
\includegraphics[width=0.475\textwidth]{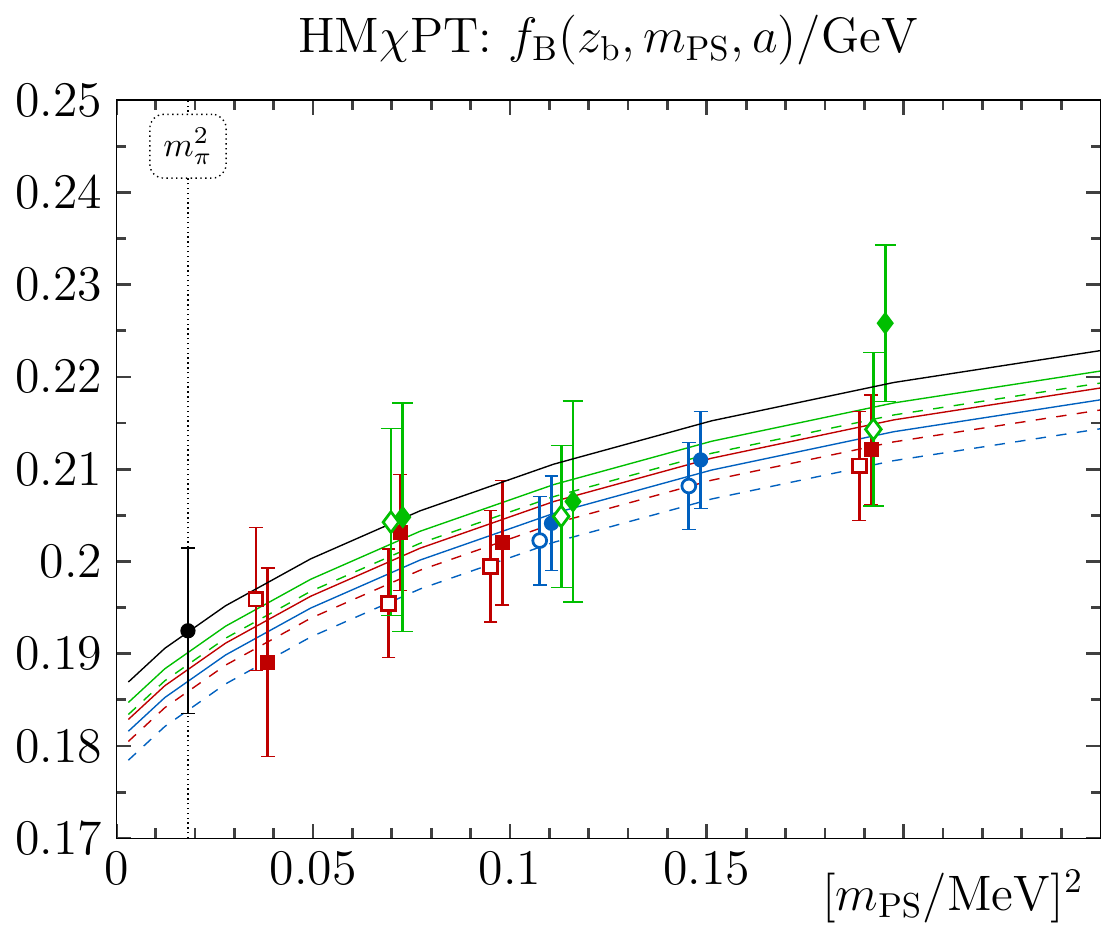}
\vspace{-0.75cm}
\caption{\sl%
Joint chiral and continuum extrapolation to the physical point of the 
B-meson decay constant (\ref{hqetexpan_fB}) in NLO HQET to the 
$\hmcpt$--motivated ansatz (\ref{fitansatzNLO_fB}).
The colour coding is the same as in Fig.~\ref{fig:Mb_LOxPT}.
(I.e., blue, red and green points refer to $\beta=5.2$, $5.3$ and $5.5$,
while filled/open symbols belong to the HYP1/2 static actions.)
}\label{fig:fB_NLOxPT}
\end{center}
\end{figure}

\begin{figure}[htb]
\begin{center}
\includegraphics[width=0.4875\textwidth]{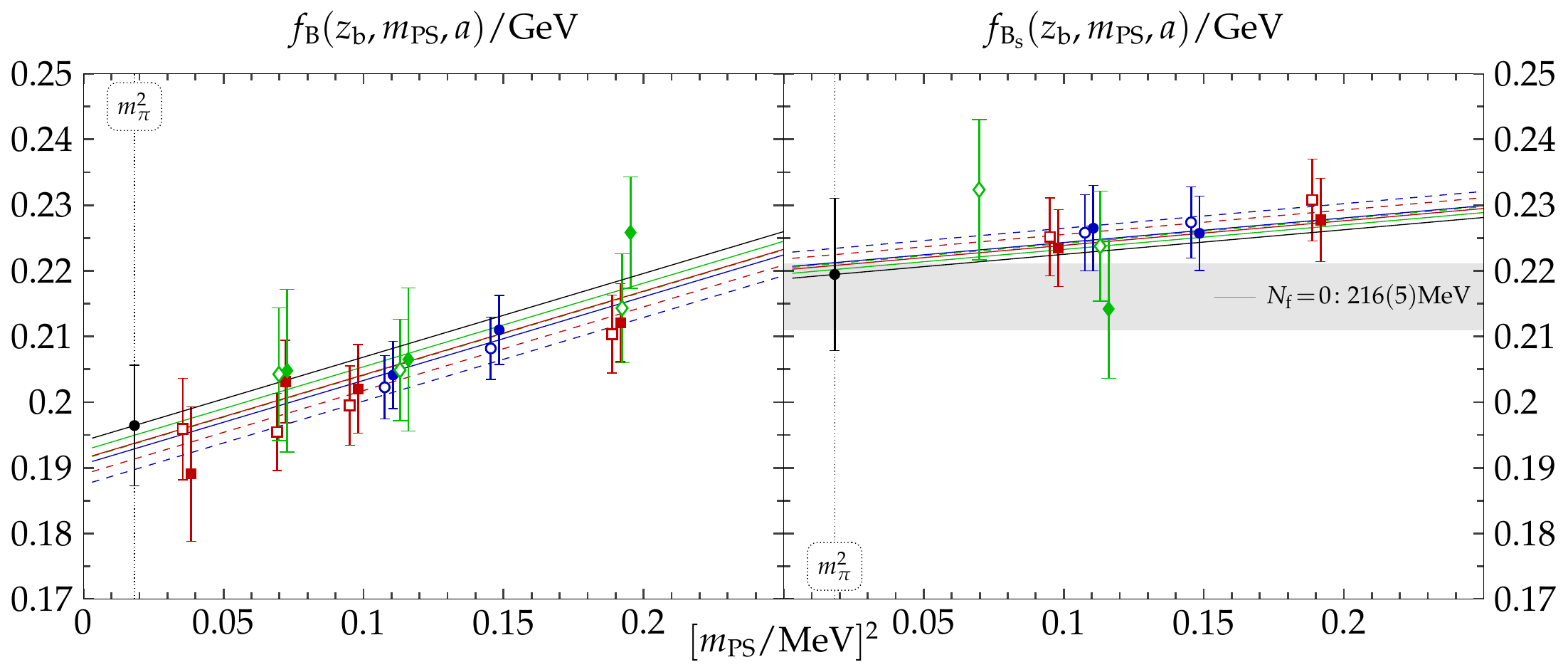}
\caption{\sl%
Joint chiral and continuum extrapolation to the physical point of $\fB$
(\textit{left}) and $\fBq{s}$ (\textit{right}) in NLO HQET, where only a 
linear dependence on the squared light pseudoscalar (sea) mass $m_{\rm PS}^2$ 
is assumed, cf.~(\ref{fitansatzLO_fB}).
In case of $\fBq{s}$, the NLO HQET result $\fBq{s}=216(5)\,\MeV$ 
obtained in the quenched approximation ($\nf=0$)~\cite{HQET:fb1m} 
(where the scale was set through $r_0=0.5\,\Fm$) is included for comparison.
The colour coding is the same as in Fig.~\ref{fig:fB_NLOxPT}.
}\label{fig:fB+fBs_LOxPT}
\end{center}
\end{figure}

Due to non-perturbative $\Or(a)$ improvement employed in our computations, 
$\fBq{r}$ (with ${\rm r}={\rm l},{\rm s}$ and $\fBq{l}\equiv\fB$), 
approaches the continuum limit quadratically in the lattice spacing.
In order to estimate a systematic error in our combined chiral and continuum 
extrapolation, we choose fits, where the sea quark dependence is modelled
according to the prediction of Heavy Meson Chiral Perturbation Theory 
($\hmcpt$)~\cite{hmchPT:G92,hmchPT:SZ96}, as well as only linear in 
$m_{\rm PS}^2$,
\bea
& &
\hspace{-1.0cm}
\fB\left(m_{\rm PS},a,{\rm HYP}n\right)=
\label{fitansatzNLO_fB}
\\
& & 
\hspace{-1.0cm}
b'\left[1-\frac{3}{4}\,\frac{1+3\,\hat{g}^2}{(4\pi f_\pi)^2}
\,m^2_{\rm PS}\ln\big(m^2_{\rm PS}\big)\right]
+c\,'m^2_{\rm PS}+d\,'_{{\rm HYP}n}\,a^2\,,
\nonumber\\[1ex]
& &
\hspace{-1.0cm}
\fBq{r}\left(m_{\rm PS},a,{\rm HYP}n\right)=  
b_{\rm r}+c_{\rm r}\,m^2_{\rm PS}+d_{{\rm r,HYP}n}\,a^2\,;
\label{fitansatzLO_fB}
\eea
here, $\fpi=f_{\pi}^{\,\rm exp}$ and $\hat{g}=0.51(2)$ are the same as above.
These joint extrapolations are depicted as the black solid curves 
in Figs.~\ref{fig:fB_NLOxPT} and \ref{fig:fB+fBs_LOxPT}.
In particular for $\fBq{s}$, not all CLS ensembles were analysed yet.
%
%
%

From the figures one can infer that whether we do or do not include the 
chiral logarithm of $\hmcpt$ in the extrapolation \eqref{fitansatzNLO_fB}
of $\fB$ induces a very small change at the physical point only.
We thus take the $\hmcpt$ extrapolation as the central value and the 
difference to the linear fit to account for a part of the systematic
error of our final result. 
For the B- and $\Bs$-meson decay constants from HQET at $\Or(1/\mb)$ in 
two-flavour QCD we preliminarily give 
\bea
\fB
& = & 
193(9)(4)_{\chi}\,\MeV\,,
\label{res_fB}
\\
\fBq{s} 
& = &
219(12)\,\MeV\,,
\label{res_fBs}
\eea
where the quoted errors again cover all sources of statistical and 
systematic uncertainties.

Our results \eqref{res_fB} and \eqref{res_fBs} are in line with computations
of other groups, see, e.g., summaries 
in~\cite{lat11:bphys,mbottom:narison_qcd12extd}. 
\section{Outlook}
\label{Sec_outl}
\noindent
The non-perturbative treatment of NLO HQET with controlled chiral and 
continuum extrapolations leads to results for B-physics phenomenology with 
a few--\% accuracy.
They can contribute to resolving current tensions in precision CKM 
analyses of the B-meson sector.
As our computations are the only ones, which have no perturbative
uncertainties, including the renormalization of the axial current, this 
introduces a new quality, albeit for $\nf=2$.  

We are also investigating further spectral quantities within our approach, 
for instance, the B-meson spin splittings.
Owing to the heavy quark spin-symmetry, the mass difference between
the vector ${\rm B^*}$- and the pseudoscalar ${\rm B}$-meson is dominated 
by a pure $\Or(1/\mb)$ effect from the contribution of $\Ospin$ to the 
effective HQET Lagrangian (\ref{lhqet}) and, hence, is of particular 
interest.

By extending our finite-volume matching strategy to all components of the 
axial and vector currents, we aim at a NLO HQET calculation of 
${\rm B}\to\pi$ semi-leptonic decay form factors as possible application.
A status report in the LO (static) approximation has been given 
in~\cite{lat12prlm:fabio}.

\section*{Acknowledgments}
\noindent
%
%
This work is supported by the DFG in the SFB/TR~9, 
``Computational Particle Physics'', and was formerly so through EU Contract 
No.~MRTN-CT-2006-035482, ``FLAVIAnet''.
We thank for further funding by the grants STFC ST/G000522/1 and 
EU ``STRONGnet'' 238353 (N.~G.), and DFG HE~4517/2-1 (P.~F. and J.~H.).
We are indebted to our colleagues in CLS for the joint production and use 
of $\nf=2$ configurations and also acknowledge the computer resources 
provided by the NIC at FZ J\"ulich, HLRN in Berlin and DESY, Zeuthen, where 
most of our simulations have been performed.
%
%
%
\bibliography{lattice_ALPHA}
\bibliographystyle{h-elsevier3.bst}
%
%
\end{document}